\documentclass[aps,prx,twocolumn,showpacs,superscriptaddress]{revtex4-1} 

\usepackage{graphicx}  % needed for figures
\usepackage{epstopdf}
\usepackage{dcolumn}   % needed for some tables
\usepackage{bm}        % for math
\usepackage{amsmath}
\usepackage{amssymb}   % for math
\usepackage{wrapfig}
\usepackage{array}
\usepackage[caption=false]{subfig}
\usepackage[bookmarks=false,linkcolor=blue,urlcolor=blue,colorlinks,citecolor=blue]{hyperref}
\usepackage{exscale,relsize}
\usepackage[usenames,dvipsnames]{color}

\makeatletter

\def\be{\begin{equation}}
\def\ee{\end{equation}}
\def\ba#1{\begin{array}{#1}}
\def\ea{\end{array}}
\def\bn{\begin{enumerate}}
\def\en{\end{enumerate}}

\def\rr{\right}
\def\l{\left}

\def\ket#1{\l|#1\rr\rangle}

\def\intt{\int\limits}

\begin{document}

\title{Robust Majorana magic gates via measurements}

\author{Torsten Karzig}

\affiliation{Microsoft Quantum, Microsoft Station Q, Santa Barbara, CA 93106-6105 USA}

\author{Yuval Oreg}

\affiliation{Department of Condensed Matter Physics, Weizmann Institute of Science,
Rehovot, 76100 Israel}

\author{Gil Refael}

\affiliation{Walter Burke Institute for Theoretical Physics and Institute for Quantum Information and Matter, California Institute of Technology, Pasadena, CA 91125 USA}
\affiliation{Department of Physics, California Institute of Technology, Pasadena, CA 91125 USA}

\author{Michael H. Freedman}

\affiliation{Microsoft Quantum, Microsoft Station Q, Santa Barbara, CA 93106-6105 USA}

\affiliation{Department of Mathematics, University of California, Santa Barbara,
CA 93106 USA}

\begin{abstract}
$\pi/8$ phase gates (magic gates or T-gates) are crucial to augment topological systems based on Majorana zero modes to full quantum universality. We present a scheme based on a combination of projective measurements and non-adiabatic evolution that effectively cancels smooth control errors when implementing phase gates in Majorana-based systems. Previous schemes based on adiabatic evolution are susceptible to problems arising from small but finite dynamical phases that are generically present in topologically unprotected gates. A measurement-only approach eliminates dynamical phases. For non-protected gates, however, forced-measurement schemes are no longer effective which leads to low success probabilities of obtaining the right succession of measurement outcomes in a measurement-only implementation. We show how to obtain a viable measurement-based scheme which dramatically increases the success probabilities by evolving the system non-adiabatically with respect to the degenerate subspace in between measurements. We outline practical applications of our scheme in recently proposed quantum computing designs based on Majorana tetrons and hexons.
\end{abstract}
\maketitle

\section{Introduction \label{sec:introduction}}

Topological quantum computation holds the promise for intrinsically error-protected storage and manipulation of quantum information using braiding of non-Abelian anyons~\cite{Kitaev03}. Majorana zero-energy modes (MZMs) form the simplest non-Abelian anyons. Formation of such a state in one- and two-dimensions was theoretically predicted to occur in quantum Hall states~\cite{Read00} and certain semiconductor-superconductor devices~\cite{Kitaev01,Sau10b,Lutchyn10,Oreg10}. In the last decade, Majorana modes have indeed emerged in reports of multiple experiments (for a recent review see Ref.~\onlinecite{Lutchyn17}), which raises the prospects for a Majorana-based topological quantum computer.

MZMs, however, are not intricate enough to permit dense population of the computational Hilbert space and, therefore, can not perform a universal topological quantum computation~\cite{Bravyi05}. While braiding of MZMs can carry out topologically protected Clifford gates, the are incapable of realizing a topologically-protected  magic gate or generating a magic state, (also known as the $T$-gate or the $\pi/8$ phase gate), which, is necessary to complete the Clifford gates, in order to realize universal quantum computation.

\begin{figure}
	\includegraphics[width=\columnwidth]{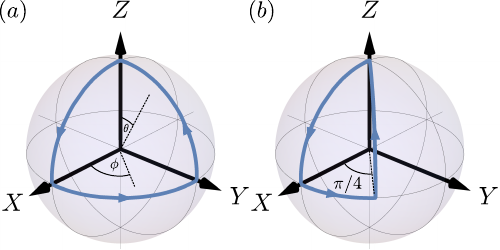}
	\caption{$(a)$ Visualization of the exchange process as a line covering an octant of the unit sphere. The line starts at the north pole ($h_z \gg h_x, h_y$), then proceeds to the $X$ point on the equator ($h_x \gg h_y,h_z$), followed by the $Y$ point ($h_y \gg h_x, h_z$) and finally reaches the north pole again ($h_z \gg h_x, h_y$), so that the cycle is completed. The Berry phase difference of the two parity sectors accumulated in this process is equal to half of the covered solid angle, $\pi/2$. $(b)$ The sequence for a $\pi/8$ gate in the ideal Y-junction system. This trajectory is not protected as we have to keep $h_x=h_y$ while modifying $h_z$, small fluctuations will yield a different phase. \label{fig:intro}}
\end{figure}

Several proposals exist for augmenting Majorana-based architectures with magic gates. These proposals, however, are generically unprotected, and range from precise timing \cite{Bravyi06,Freedman06,Sau10c,Hyart13}, over topological/non-topological hybrid systems \cite{Jiang11a,Bonderson11,Hassler11,Hoffman16}, to fine-tuned geometric approaches~\cite{Clarke16,Plugge16,Plugge17}. An exception is a proposal to develop highly specialized genon-based  hardware to produce a topologically protected magic-gate~\cite{Barkeshli13a, Barkeshli15}. In contrast, the current authors proposed a geometric protocol which is robust against systematic errors~\cite{Karzig16}. From a geometric point of view, a $\pi/4$ phase gate that corresponds to an exchange of two MZMs corresponds to a topologically protected adiabatic path encircling an octant in the Bloch sphere of the parameter space, see Fig.~\ref{fig:intro}a. By traversing the geometric space in an alternating way it is possible to implement a geometric decoupling scheme for a $\pi/8$ phase gate which effectively cancels systematic errors so that the remaining error is exponentially small in the number of turns that are properly chosen at zeros of Chebyshev polynomials~\cite{Karzig16}. As we review in Sec.~\ref{sec:evolution}, however, unfortunately and unavoidably, leaving the protected path defined by the edge of the octant, the $\pi/8$ gate cannot be realized in a fully geometric way since the system will, in general, pick up a small but finite dynamical phase that has to be eliminated with conventional error correcting protocols (e.g., through echo sequences).

In the current work, we propose using elements of measurement-only topological quantum computation~\cite{Bonderson08b} to overcome these dynamical errors. We demonstrate a sequence of protocols, starting with a simple measurement-based realization of the geometric magic gate of Ref. \cite{Karzig16}, then modifying this sequence with additional measurements, and, finally, combining dynamical evolution and measurements to yield a superior protocol that eliminates systematically all leading sources of error. 

On general grounds, one might expect that the measurement-only schemes can perform better than the adiabatic braiding methods as part of the information (the results of the projective measurements) are stored and used classically, and, hence, do not experience any quantum de-coherence effects. To illustrate that, consider four MZMs operators ($\gamma_0, \gamma_x, \gamma_y, \gamma_z $). Using the relation $\left\{\gamma_i, \gamma_j \right\} =2\delta_{ij},\; i,j=0,x,y,z$. It can be easily checked that with the help of ancillas $\gamma_0,\gamma_z$ the braiding operator of $\gamma_x$ and $\gamma_y$ given by ${\cal B}_{xy}=e^{\frac{\pi}{4} \gamma_x \gamma_y}$ can be realized by a sequence of projections $P_z P_y P_x P_z={\cal B}_{xy}P_z/\sqrt{8}$ with $P_i=(1-i\gamma_0 \gamma_i)/2$ being the projective measurement operator on the states where the MZM pair $\gamma_0$ and $\gamma_i$ is unoccupied. Since in a typical measurement, the probability of measuring even parity of a pair of MZM, i.e., projecting the pair onto the unoccupied state, equals $1/2$, the total chance for applying the above projectors equals $1/2^3=1/8$. Obtaining other measurement outcomes will, in general, create a different gate which might require suitable corrections depending on the outcomes. An alternative is to use a forced-measurement scheme in which a pair of measurement processes is repeated until the desired measurement outcomes are obtained~\cite{Bonderson08b}.

The structure of the manuscript is as follows. In Sec.~\ref{sec:old_scheme1} we translate the adiabatic geometric decoupling scheme to a measurement-only procedure. We suggest applying successive projection operators at the turning points of the geometric decoupling trajectory of the adiabatic scheme. The measurement-only scheme allows us to avoid the situation where all Majorana couplings are significant, which is where dynamical phases are accrued. While this removes the need for an echo error correction, the success probability in this scheme becomes dependent on the state of the qubit, and lead to small deviations from the desired phase gate. In Sec.~\ref{sec:old_scheme2}, we show how these deviations could be avoided through a forced-measurement echo procedure, reminiscent of the dynamical phase cancellation echo for the adiabatic scheme. In Sec.~\ref{sec:ns_measurement} we discuss a north/south projection protocol that completely eliminates the need for echo procedures, and renders an accurate magic gate, but with a success probability fall off as $2^{-N}$, where $N$ is the number of steps in the geometric decoupling protocol. Since only successful outcomes are fed into a subsequent distillation scheme small success probabilities are, in principle, not problematic. Increasing the success probability, however, will drastically reduce the time to prepare a magic state. 

The main result in this manuscript is a protocol that combines dynamical evolution with measurement steps. In Sec.~\ref{sec:hybrid} we show that a novel hybrid evolution/measurement approach raises the success probability to ${\cal O}(1)$, while also producing an accurate high-quality phase gate

In Sec.~\ref{sec:remaining}  we discuss the remaining sources of errors of the hybrid scheme. The first is due to dissipation acting perpendicular to the applied Hamiltonian. The second is due to fast temporal noise affecting the control of the system. Following, Sec.~\ref{sec:numerics} is devoted to numerical implementation of the hybrid evolution exemplifying the various methods discussed in Secs. \ref{sec:Measurement-only} and \ref{sec:hybrid}. Physical realizations of the proposed scheme in the hexon and tetron geometries are discussed in Sec.~\ref{sec:realization}. Finally, in Sec.~\ref{sec:Conclusions} we summarize, conclude and discuss future prospects of the results of this study.

To focus on the non-protected gate operations, we assume throughout the paper that the protection of topological operations is perfect and thus neglect exponential small corrections (including finite-size and finite-temperature effects) as well as quasi-particle poisoning. 

\section{Review of geometric decoupling} \label{sec:evolution}

The main problem of realizing a robust magic gate is the need for extreme fine tuning of the qubit Hamiltonian. Despite their topological protection, MZMs are no exception. MZMs, however, offer a relative advantage over non-topological qubits since it is possible to exploit geometric phases. Below we recall the geometric procedure for obtaining a magic state using MZMs, consider the leading pitfalls of the procedure, and show how to use a universal geometric-decoupling procedure to overcome the bulk of the errors \cite{Karzig16}.

\begin{figure}
	\includegraphics[width=0.6\columnwidth]{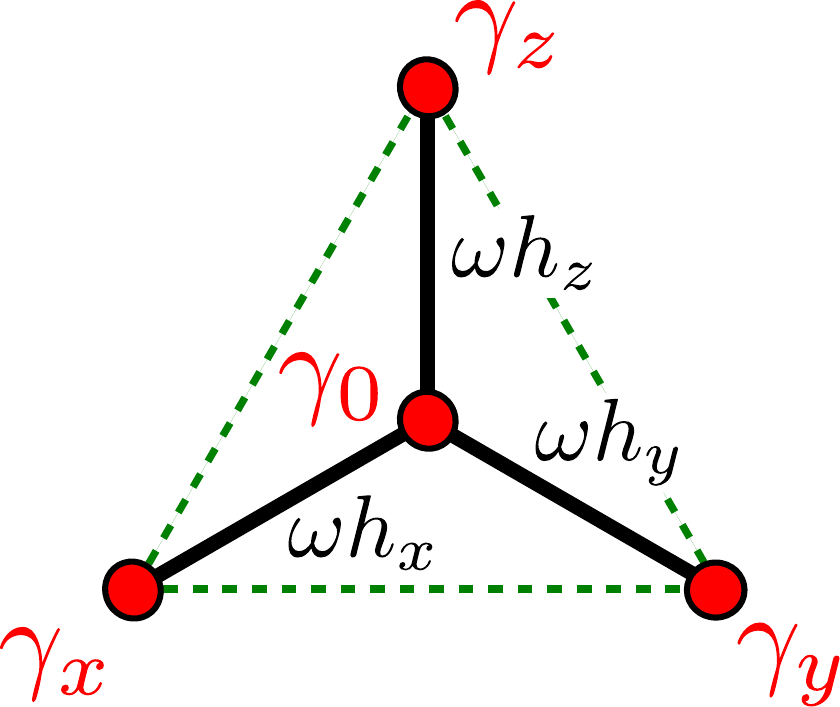}
	\caption{Schematic figure of a Y-junction corresponding to Hamiltonian \eqref{eq:Ham}. Direct coupling terms $\omega h_i$ between Majorana modes are denoted by thick black lines, while possible next-neighbor (outer) couplings are denoted by dashed green lines.}
	\label{fig:Y_juntion}
\end{figure}

The geometric path to a magic gate is best illustrated with a Y-junction system (see Fig.~\ref{fig:Y_juntion}). Three Majorana modes, $\gamma_{x},\,\gamma_y$ and $\gamma_z$, are located at the tips of the Y-junction and interact only with the fourth MZM, $\gamma_0$, which is at the center of the junction, with Hamiltonian:
\be
H=i\frac{\omega}{2}\gamma_0(\boldsymbol{h}\cdot \boldsymbol{\gamma}),
\label{eq:Ham}
\ee
where we conveniently defined the Majorana vector $\boldsymbol{\gamma}=(\gamma_x,\,\gamma_y,\,\gamma_z)$ and the coupling unit vector $\boldsymbol{h}=(h_x,\,h_y,\,h_z)$. The Y-junction couplings, $\omega h_i$, depend exponentially on physical parameters, such as the distance between the MZMs, or gate-controlled confining potentials (see e.g. \cite{Heck12,Karzig17}). This motivates the fundamental assumption in this paper that these couplings can be tuned so that their ratios reach $0$ or $\infty$ with exponential accuracy.

\subsection{Exchange process and its $\pi/8$ (magic) generalization}

An exchange process of MZMs in this system can be implemented by tuning the strengths of the couplings $\omega h_i$. Start with $h_z \approx 1 \gg h_{x}, h_{y}$. $\gamma_x$ and $\gamma_y$ are then zero modes of the problem. To exchange them, move to $h_x \approx 1 \gg h_{z}, h_{y}$ in a continuous fashion while keeping $h_y \ll 1$. Followed by $h_y \approx 1 \gg h_x,h_z$ (while keeping $h_z \ll 1$) and finally returning the system to its original state $h_z \gg h_x,h_y$ (while keeping $h_x\ll 1$).

Such manipulations can be geometrically visualized.  Let us think of $\boldsymbol{h}$ as a 3D vector, and represent it with spherical coordinates \cite{Chiu15}. $\boldsymbol{h}$ is then the radius-vector, and we also use the polar and azimuthal angles $\theta$ and $\phi$ and their unit vectors $\boldsymbol{e}_{\theta}$ and $\boldsymbol{e}_{\phi}$. With this, the Majorana states
\be
\gamma_{\theta}=\boldsymbol{\gamma}\cdot \boldsymbol{e}_{\theta},\,
\gamma_{\phi}=\boldsymbol{\gamma}\cdot \boldsymbol{e}_{\phi},
\label{eq:gamma}
\ee
are zero modes which commute with the Hamiltonian~(\ref{eq:Ham}).
The exchange process is now easily visualized as $\boldsymbol{h}$ marking a unit-sphere octant, bounded between the $\phi=0$, $\theta=\pi/2$ and $\phi=\pi/2$ planes (see Fig.~\ref{fig:intro}a).

The effect of such adiabatic manipulations on the state of the two zero modes is encapsulated in the Bloch sphere Berry phase that the $\boldsymbol{h}$ demarcates.  Writing a single Fermi annihilation operator from the two zero modes as
\be
\label{eq:a}
a=\frac{1}{2}\l(\gamma_{\theta}+i\gamma_{\phi}\rr).
\ee
This operator connects two parity states, $\ket{0}$ (defined by $a\ket{0}=0$, and $a^{\dagger}\ket{0}=\ket{1}$). Upon adiabatic closed manipulation of the vector $\boldsymbol{h}$, these states change as:
\begin{equation}
U_c\ket{(1\pm 1)/2}=e^{\pm i\alpha}\ket{(1\pm 1)/2},
\end{equation}
where the phase difference $2\alpha$ is given by the solid angle enclosed by the demarcated contour. For an octant, we indeed, obtain $\alpha_{\rm exchage}=\pi/4$.

Obtaining the magic $\pi/8$ gate now seems palpable. All we need is to cover half the solid angle that the exchange process covers. For instance, we could start with $\theta=\phi=0$, turn $\theta=0\rightarrow\pi/2$, then $\phi=0\rightarrow \pi/4$, and return with $\theta=\pi/2\rightarrow 0$. Finally, $\phi=\pi/4\rightarrow 0$ closes the trajectory (Fig.~\ref{fig:intro}b).

Despite the elegance of the geometric magic gate, it suffers a crucial pitfall. The $\phi=\pi/4$ plane is a fine tuned swath of parameter space, which requires keeping $h_x=h_y$. But such control is not realistic, and control errors will lead to an arbitrary error in the computation.  Additionally, and crucially for our current work, the $\pi/8$ trajectory must go through a region where all three Majorana couplings have similar strengths ($\theta=\phi=\pi/4$). This will unavoidably induce next-nearest-neighbor couplings (see Fig.~\ref{fig:Y_juntion}) between the Majorana modes at the Y-junction tip, which will split the ground-state degeneracy between the two parity states, and induce an arbitrary dynamical phase between the $\ket{0}$ and $\ket{1}$ states. In the following, we refer to direct coupling between the MZMs, $\gamma_x,\gamma_y$,and $\gamma_z$ as outer couplings.

\subsection{Universal geometric decoupling \label{sec:GD}}

\begin{figure}
\includegraphics[width=0.8\columnwidth]{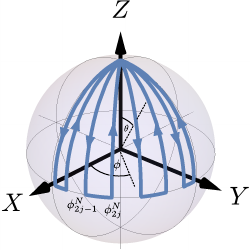}
\caption{Evolution-based geometric decoupling scheme. A proper choice of the turning point $\phi_n^N$ yields a trajectory covering a solid angle of $\pi/4$ with an exponentially small error. Here we plot the contour for the Chebyshev polynomials with $N=5$ and $\phi_n^N, n=1,\dots, 2N$ are given in Eq.~(\ref{eq:phinNchebyshev})}
\label{fig:old_scheme}
\end{figure}

The crux of Ref.~\cite{Karzig16} is to show that the systematic control error described above could be eliminated to arbitrary precision using an iterative and universal trajectory through the $\boldsymbol{h}$ sphere.
The idea follows from the intuition that snake-like trajectories as in Fig.~\ref{fig:old_scheme} could essentially average out the error that arises from the imperfect control of the device. The turning points $\phi_1^N,\phi_2^N,\dots,\phi_n^N,\;\;\; n=1,\dots,2N$ can be optimized to systematically eliminate the error in the accumulated phase reminiscent to the concept of universal dynamical decoupling \cite{Uhrig07}.

In particular, is becomes possible to exponentially suppress gate errors in the number of turns $\delta \alpha\sim {\rm e}^{-2N}$, as long as the errors in the turning points $\delta \phi_n^N$ are systematic and described by a smooth function $\delta \phi_n^N=\delta \phi (\phi_n^N)$. The optimal turning points can then be derived by expanding the errors in terms of Chebyshev polynomials and eliminating the first $2N-1$ orders of the expansion~\cite{Karzig16}. This procedure yields $2N$ equations
\begin{equation}
\sum_{n=1}^{2N}(-1)^{n} T_m^*\left(\frac{2}{\pi}\phi_n^N\right)=\frac{4}{\pi}\alpha\left(1-(-1)^m\right), \label{eq:phin}
\end{equation}
with $m=1\dots 2N$, where $T_m^*(x)=T_m(2x-1)$ are shifted Chebyshev polynomials of the first kind. A magic gate is implemented for $\alpha=\pi/8$. In that case, the solutions $\phi_n^N$ can be expressed analytically and are given by
\begin{equation}
\label{eq:phinNchebyshev}
\phi^N_{n}=\frac{\pi}{4}\left[1-\cos\left(\frac{\pi n}{2N+1}\right)\right].
\end{equation}

The Chebyshev protocol, while efficiently eliminating the systematic machine error, does not solve the problem of the uncontrollable dynamical phase due to finite outer couplings that arise when all the couplings $h_i$ are strong.  In Ref. \cite{Karzig16} we illustrate how this dynamical phase can be eliminated by carrying out an echo sequence. Echo sequences, however, could prove costly, as they lengthen the calculation time, and rely strongly on the stability of the system. Our current work seeks to eliminate the need for an echo altogether by avoiding the regions where all three couplings $\omega h_i$ are sizable. This can be done using a measurement-based approach as we show below.

\section{Measurement-only approach}
\label{sec:Measurement-only}

In a measurement-only approach, the adiabatic evolution of the state is replaced by applying a set of measurements \cite{Bonderson08b}. Here, we focus on measurements that determine the (joint) parity of a set of MZMs. With the knowledge of a measurement outcome a measurement can be described by projectors $P_p$ or $P_{\bar{p}}$, where $p$ denotes the parity of the selected set of MZMs and we define $P_p\,(P_{\bar{p}} )$ as the projection onto $p=1\, (p=-1)$. A series of measurements then acts on an initial state $|\psi\rangle$ as a product of projectors yielding the (normalized) final state $p_\text{s}^{-1/2}\prod_j P_j |\psi\rangle$ where $p_\text{s}$ denotes the probability of obtaining the specific set of measurement outcomes.

\subsection{Direct translation of evolution-based to measurement-only geometric decoupling}
\label{sec:old_scheme1}

\begin{figure}
	\includegraphics[width=\columnwidth]{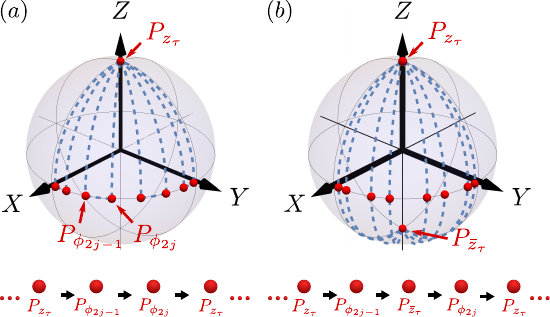}
	\caption{Measurement-only geometric decoupling schemes. The projection operators are applied in order indicated below each panel. $(a)$ Direct translation of the evolution based scheme in Fig.~\ref{fig:old_scheme} to a measurement-only implementation. $(b)$ Measurement-only implementation of the north/south sweep protocol.}
	\label{fig:measurement_only}
\end{figure}

Let's now return to the 4 MZM system discussed in Sec.~\ref{sec:evolution}. The evolution-based geometric decoupling scheme can be translated into a measurement-only protocol by applying projectors at all the turning points of the adiabatic evolution along the Bloch sphere (see Fig.~\ref{fig:measurement_only}). In particular, the evolution from the north pole to equator at azimuthal angle $\phi_{2j-1}$, along the equator to $\phi_{2j}$, and back to the north pole is described by a set of projectors $\Pi_{\phi_{2j}-\phi_{2j-1}}\equiv P_{z_\tau} P_{\phi_{2j}} P_{\phi_{2j-1}}P_{z_\tau}$, with
\begin{eqnarray}
2P_{z_\tau} &=&1-i\gamma_0\gamma_z \\
2P_\phi &=&1-i\gamma_0\big[\cos(\phi)\gamma_x+\sin(\phi) \gamma_y\big]\,.
\end{eqnarray}

In the following, it will be convenient to rewrite pairs of MZMs in terms of Pauli matrices $\sigma_j$ as $i\gamma_0 \gamma_x=\sigma_x$, $i\gamma_0\gamma_y=\sigma_y$ and $-i\gamma_x \gamma_y=\sigma_z$. Note that $i\gamma_0\gamma_z=\sigma_z\tau_z$ where the Pauli matrix $\tau_z=-\gamma_x\gamma_y\gamma_z\gamma_0$ describes the overall parity of the 4 MZM system. For concreteness, we focus in the following on the case of a system with 6 MZMs at a fixed (even) parity where a qubit is encoded in the 4 MZMs $\gamma_x,\gamma_y, \gamma_1, \gamma_2$ with $\gamma_0,\gamma_z$ acting as ancillas. In this case $\tau_z=i\gamma_1\gamma_2$ describes the $z$-Pauli operator of the qubit.

In this notation we find
\begin{eqnarray}
 \Pi_{\Delta \phi}&=&\frac{1}{4}P_{z_{\tau}}\left(1+\cos(\Delta \phi)-i\sin(\Delta\phi)\sigma_z\right) \\
 &=&\frac{1}{2}\cos(\Delta \phi/2)e^{i\tau_z\Delta\phi/2}P_{z_\tau}\,, \label{eq:Pi_delphi}
\end{eqnarray}
where we used that $P_{z_\tau}$ projects onto the space where $\sigma_z=-\tau_z$. Equation~\eqref{eq:Pi_delphi} allows to group the effect of the set of projectors  $\Pi_{\Delta \phi}$ into three different contributions: (1) The projector $P_{z_\tau}$ fixes the state of the ancillary ($\sigma$) degree of freedom. (2) The unitary operator $\exp(i\tau_z\Delta\phi/2)$ acts as a phase gate on the qubit ($\tau$) degree of freedom. The origin of the phase gate stems from picking up different geometric phases when completing a cyclic path in the ancillary ($\sigma$) Bloch sphere. Depending on whether $\tau_z=+1(-1)$ the path starts and ends at the north (south) pole. The opposite geometric phases $\pm \Delta \phi/2$ of the two cases then act as a phase gate. (3) The prefactor describes the success probability  $\cos^2(\Delta \phi/2)/4$ of obtaining the measurement results.

Note that the phase gate that is implemented by $\Pi_{\Delta\phi}$ is the same as that of the adiabatic evolution along the geodesics connecting the projection points on the Bloch sphere. This correspondence allows to implement the same geometric decoupling scheme of Sec.~\ref{sec:evolution} in a measurement-only setting. The full $\pi/8$ gate would then be implemented by $\prod_{j=1}^N \Pi_{\phi_{2j}-\phi_{2j-1}}$. There is, however, an important difference: only a specific set of measurement outcomes yields the individual projectors~$\Pi_{\phi_{2j}-\phi_{2j-1}}$.

Let's now study the resulting operation for different measurement outcomes. The effect of flipping $P_{\phi}\rightarrow P_{\bar{\phi}} = P_{\phi+\pi}$ is relatively minor. If both angles are shifted by $\pi$ we still obtain the same gate. If only one angle is shifted the gate differs by an overall $\tau_z$ gate which can easily be recorded and (if needed) corrected. Problems would arise if the $z$ projections shift $P_{z_\tau}\rightarrow P_{\bar{z}_\tau}$. The latter would lead to random sign flips $\exp(\pm i\tau_z \Delta \phi /2)$. To avoid this issue one could used forced measurements~\cite{Bonderson08b} by repeating the the measurements along the axis of $\phi_{i}$ and $z_\tau$ until we find $z_\tau=+1$. Note that different paths in the course of this correction procedure only lead to phase differences that are multiples of $2\pi$ which can be ignored~\footnote{Compare, e.g., the geometric phases of the path through the points $(z_\tau,\phi)=(0,\phi_{2j-1})\rightarrow (+1,\phi_{2j-1})$ with $(0,\phi_{2j-1})\rightarrow (-1,\phi_{2j-1}) \rightarrow (0,\phi_{2j-1}+\pi)\rightarrow (+1,\phi_{2j-1})$}. The forced-measurement procedure, therefore, allows to increase the success probability of the entire measurement protocol from $2^{-N}$ to unity.

\subsection{Effect of outer couplings}
\label{sec:old_scheme2}

One of the motivations of employing a measurement-based scheme to implement geometric decoupling is that by using projections instead of adiabatic time evolution it is possible to avoid the middle region of the octant where all $\omega h_i$ Majorana couplings are turned on. As mentioned in Sec.~\ref{sec:evolution} the danger of that regime is that unavoidable second-order couplings between ``outer'' MZMs (e.g. between $\gamma_x$ and $\gamma_y$) lead to a splitting of the degeneracy of the qubit states. The accompanying dynamical phases then have to be canceled by an extra echo procedure.

Dynamical phases do not appear when applying projectors of the type of Eq.~\eqref{eq:Pi_delphi} since there is always at least one MZMs that is untouched and therefore guaranties a perfect ground state degeneracy. The outer couplings, however, can lead to a different error source by applying modified projectors
\begin{eqnarray}
2P_{\phi,\vartheta} &=&1-i\gamma_0\cos(\vartheta)\big[\cos(\phi)\gamma_x+\sin(\phi) \gamma_y\big] \nonumber \\ 
&& -i\sin(\vartheta)\gamma_x\gamma_y
\label{eq:Pphitheta}
\end{eqnarray}
instead of $P_\phi=P_{\phi,\vartheta=0}$. 

This form of projectors emerges when considering how to physically implement the corresponding measurements. A measurement of MZM parities can be thought of as a two step process. First, the four-fold ground state degeneracy is split (once) by introducing couplings between MZMs either internally or through coupling to the measurement apparatus. This process is described by a Hamiltonian $H_M$. The finite energy splitting then allows the measurement apparatus to determine whether the system is in the ground or excited state by an appropriate energy spectroscopy. We will discuss details of the concrete implementation of measurements in hexon and tetron geometries in Sec.~\ref{sec:realization}. In the following, we will describe a measurement by turning on a Hamiltonian $H_M$ and then projecting onto the corresponding energy eigenstates.

Measurements of the form \eqref{eq:Pphitheta} are implemented by the Hamiltonian
\begin{equation}
H_M=i\frac{\omega}{2}[\cos(\vartheta)(\cos(\phi)\gamma_0\gamma_x+\sin(\phi)\gamma_0\gamma_y)+\sin(\vartheta)\gamma_x\gamma_y]\,,\label{eq:outer}
\end{equation}
where $\tan(\vartheta)$ quantifies the ratio of the outer to inner couplings. While it is in principle possible to fine-tune to $\vartheta=0$, in general one expects a second order coupling $\sin(\vartheta)\sim \cos(\phi)\sin(\phi)\omega/\Delta_0$ where $\Delta_0$ is a higher energy scale (e.g. the topological gap) that was integrated out to obtain the effective MZM Hamiltonian $H_M$.

The effect of a finite $\vartheta$ is that $P_{\phi,\vartheta}$ no longer projects directly onto the equator of the ($\sigma_x,\sigma_y,\sigma_z\tau_z$) Bloch sphere, but rather onto points shifted a little to the north or south. Interestingly, since these shifts are opposite for different states of the $\tau_z$ qubit, they do not affect the geometric phase that is picked up when applying $\Pi_{\delta \phi, \{\vartheta\}}=P_{z_\tau}P_{\phi_{2j},\vartheta_{2j}}P_{\phi_{2j-1},\vartheta_{2j-1}}P_{z_\tau}$. In particular, to linear order in $\{\vartheta_{2j},\vartheta_{2j-1}\}$ we find (up to an overall phase)
\begin{equation}
\Pi_{\delta \phi,\{\vartheta\}}=\left(1-\bar{\vartheta}\tau_z\right)\Pi_{\delta \phi}\,,
\label{eq:Pi_phitheta}
\end{equation}
with $\bar{\vartheta}=(\vartheta_{2j}+\vartheta_{2j-1})/2$. While a finite $\bar{\vartheta}$ does not change the applied phase rotation, the (real) prefactor now becomes $\tau_z$ dependent. The latter follows intuitively from different success probabilities of the projections since depending on the state of the $\tau_z$ qubit, the projection $P_{\phi,\vartheta}$ is either closer to or further away from the north pole. Unfortunately, due to the presence of a $\tau_z$ dependent prefactor, projections of the type of Eq.~\eqref{eq:Pi_phitheta} can no longer be used to prepare precise magic states \footnote{Since $X$ eigenstates \unexpanded{$(|0\rangle +|1\rangle)/\sqrt{2}$} can be prepared with topological accuracy a precise way of preparing a magic state is applying a $\pi/8$ phase gate to initial $X$ states. Applying a gate of the form $(1-\bar{\vartheta}\tau_z)\exp(i\tau_z\pi/8)$ would rotate $X$ states out of the equator and introduce errors.}.

Note that an additional echo similar to the one that cancels the dynamical phases in the evolution-based scheme could be used to cancel the $\tau_z$ dependent prefactor: First, apply the geometric decoupling protocol to implement a $\pi/16$ phase gate with some unwanted overall prefactor $(1-\bar{\vartheta}\tau_z)$. Then, use the forced measurement scheme to project on the south instead of the north pole and reverse the order of the turning points to still implement a $\pi /16$ gate (this step is equivalent to flipping the qubit and applying a $-\pi/16$ gate). The result will be a $\pi/8$ gate where the $\tau_z$ dependence in the prefactor was eliminated via $(1-\bar{\vartheta}\tau_z)(1+\bar{\vartheta}\tau_z)=1-\bar{\vartheta}^2$ for each turn.

\subsection{North/south projection protocol}
\label{sec:ns_measurement}

Within the protocols discussed in Secs.~\ref{sec:old_scheme1} and~\ref{sec:old_scheme2},  the evolution and measurement-only approaches are very similar, with essentially a one-to-one mapping of their strengths (eliminating systematic errors) and weaknesses (requiring some sort of echo procedure). We now present a different protocol that allows a measurement-only scheme to eliminate the effect of unwanted outer couplings without the need for extra echos. 

The minimal building block of the protocol is given by $\Pi^{\text{ns}}_{\Delta \phi,\{\vartheta\}}=P_{z_\tau}P_{\phi_{2j},\vartheta_{2j}}P_{\bar{z}_\tau}P_{\phi_{2j-1},\vartheta_{2j-1}}P_{z_\tau}$ and describes projections in the Bloch sphere from north to south and back (see Fig.~\ref{fig:measurement_only}b). Since any projector $P_{\phi,\vartheta}$ is enclosed by antipodal projections in $Z$ direction only terms that do not commute with $\sigma_z\tau_z$ survive which cancels the unwanted terms $i\gamma_x\gamma_y=-\sigma_z$. The resulting projection yields
\begin{eqnarray}
\Pi^\text{ns}_{\Delta \phi,\{\vartheta\}}&\! =\!&\frac{1}{4}\!\cos(\vartheta_1)\!\cos(\vartheta_2)P_{z_{\tau}}\!\sigma_x e^{i\sigma_z\phi_2}\!P_{\bar{z}_{\tau}}\!\sigma_x e^{i\sigma_z\phi_1}\!P_{z_{\tau}}\ \ \ \ \   \\
&\! =\!&\frac{1}{4}\cos(\vartheta_1)\!\cos(\vartheta_2)e^{i\tau_z \Delta \phi}P_{z_{\tau}}\,
\label{eq:ns_projector}
\end{eqnarray}
with $\Delta \phi=\phi_2-\phi_1$. The implemented gate has the same form as the version with $\vartheta=0$ of Eq.~(\ref{eq:Pi_delphi}), with the difference that the accumulated phase is now doubled. We can, therefore, use a succession of projections $\Pi^\text{ns}_{\Delta \phi,\{\vartheta\}}$ to implement the full geometric decoupling scheme when choosing turning points $\phi_i$ suitable for a $\pi/16$ gate in the original protocol. I.e., using solutions of Eq.~\eqref{eq:phin} with $\alpha=\pi/16$. Conceptually, a similar procedure would also be possible in an adiabatic-evolution based scheme. The requirements for the control of the Hamiltonian to achieve the desired cancellation of dynamical phases are, however, much harder to meet. One would need to change the sign of the $Z$ component of the Hamiltonian while keeping the $X$ and $Y$ parts exactly the same as for the evolution through the northern hemisphere. For the measurement-only version, projecting onto the north or south pole corresponds to applying exactly the same measurements; we simply have to select for different measurement outcomes.

As we see from Eq.~\eqref{eq:ns_projector}, the effect of finite angles $\vartheta_i$ is minimal, as it simply slightly reduces the success probability of the set of measurement outcomes. Similar to Sec.~\ref{sec:old_scheme1}, the protocol works with any outcome of the measurements along the equator. Here, we don't even need to record the outcomes since the corresponding paths only differ by great circles, and thus by phases of $2\pi$. Measurement outcomes, however, do matter for measurements along the $Z$ axis and completing the progression from north $\rightarrow$ south $\rightarrow$ north has a likelihood of $1/4$ (for $\vartheta=0$).

Obtaining the wrong measurement outcomes leads to contributions $P_{z_\tau}P_{\phi,\vartheta}P_{z_\tau}=(\cos(\vartheta)-\sin(\vartheta)\tau_z)P_{z_\tau}/2$ which reintroduce the unwanted $\tau_z$ dependent prefactors. This prevents an efficient implementation of a forced-measurement scheme since the $\tau_z$ dependent terms of wrong measurement outcomes would need to be appropriately canceled. Nevertheless, with a probability $2^{-2N}$ the protocol yields an implementation of the geometric decoupling scheme that does not suffer from the effects of unwanted couplings and with the recorded outcomes of the measurement along the $Z$ axis it is also known when this scenario was realized. In the next section we discuss how to drastically increase the probability of obtaining the right measurement outcomes by utilizing a hybrid evolution-measurement scheme.

\begin{figure}
	\includegraphics[width=0.8\columnwidth]{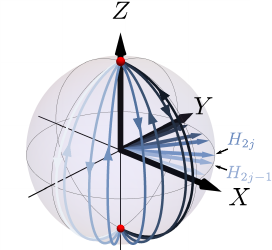}
	\caption{Visualization of the hybrid protocol. The lines indicate the evolution of the state while the arrows of the same color indicate the corresponding Hamiltonian terms $H_{2j-1}$ and $H_{2j}$ that drive the precession. For illustrative purposes we showed the back-evolution from the south to the north pole shifted by $\pi$ corresponding to applying $-H_{2j}$. The sign of $H$ does not matter since it leads to phase differences of $2\pi$. At the end of each evolution stage a projection, with high success probability, is performed to the south or north pole as indicated by the red dots.}
	\label{fig:ns}
\end{figure}

\section{Hybrid evolution and measurement scheme}
\label{sec:hybrid}

By combining measurements with dynamical evolution of the system we can obviate the problem of low success probabilities of the measurement-only north/south projection scheme. In this proposed procedure, the projective measurements along the north and south poles are supplemented by a free (non-adiabatic) evolution which shifts the state of the four MZMs between the two poles. By combining the measurement with the free evolution we manage to be dramatically reduce the probability of measurement errors, as well as eliminate the possibility of static machine error (e.g., the $\vartheta\neq 0$ case from Eq. \eqref{eq:Pphitheta}).

The proposed procedure is as follows. Just as in the measurement-only procedure we break the evolution into a series of wedges flowing from the north pole to the south pole and back through the azimuthal angles $\phi_{2j-1}$ and $\phi_{2j}$, respectively. The procedure for the $j$th wedge begins by projecting the qubit into its north pole, $P_{z_{\tau}}$. Next, ideally, we evolve the qubit with the Hamiltonian:
\begin{eqnarray}
H_j &=&\frac{\omega}{2} i\gamma_0(\gamma_x\cos \phi_j+\gamma_y \sin\phi_j)\vspace{2mm} \nonumber\\
&=&\frac{\omega}{2} \boldsymbol{h}_{\phi_j}\cdot \boldsymbol{\sigma}\label{rot}\,
\end{eqnarray}
with $\boldsymbol{h}_{\phi_j}=\{\cos(\phi_j),\sin(\phi_j),0\}$ and $\boldsymbol{\sigma}=\{\sigma_x,\sigma_y,\sigma_z\}$.
Note that in contrast to adiabatic evolution schemes the system starts out with all couplings turned off after the projection to then north pole. Turning on the above Hamiltonian is, therefore, necessarily non-adiabatic and acts like a perpendicular magnetic field in which the ``spin'' of the qubit Bloch sphere freely precesses. This precession will evolve the state from the north pole along the great arch intersecting the equator at $\phi_j-\pi/2$ until it reaches the south pole where $P_{\bar{z}_{\tau}}$ should be measured to be 1. For this purpose, we turn the Hamiltonian (\ref{rot}) on for a time $T=\pi(1+2n)/\omega$. It may appear as though this time has to be fine tuned, but that is not the case. The measurement that follows will align the qubit with south pole precisely which corrects over or under rotation. As we show below, the main virtue of a precise adjustment of the free precession time is an increase of the success probability of the process. After the projection to the south pole we apply the Hamiltonian
\be
\ba{c}
H_{j+1}=-\frac{\omega}{2} \l(\cos\phi_{j+1} \sigma_x+\sin\phi_{j+1}\sigma_y\rr)\label{rot1}
\ea
\ee
for a time $T=\pi(1+2n')/\omega$. This will return the qubit to the vicinity of the north pole, where a measurement in the $Z$ basis will lead to a likely projection onto $P_{z_{\tau}}$. Note that the minus sign in Eq.~\eqref{rot1} is not necessary and was chosen for convenience to align with the wedges in Fig.~\ref{fig:ns}. Shifting $\omega \to -\omega$ leads to an essentially equivalent path that differs by a phase of $2\pi$.

Let  us next calculate the evolution of the qubit state during the above one-wedge process. The operator applied to the qubit state reads, in terms of the Pauli matrices:
\be
\Pi^\text{hyb}=P_{z_{\tau}}e^{-iTH_{j+1}}P_{\bar{z}_{\tau}}  e^{-iTH_{j}}P_{z_{\tau}}
\ee
At this point, we assume that the Hamiltonian $H_{j}$ is not perfect, and has a small component of $i\gamma_x\gamma_y=\sigma_z$ mixed in it, due to unavoidable outer couplings as in Eq.~\eqref{eq:outer}:
\be
H_j=\frac{\omega}{2}\l[\cos\vartheta_j \l(\cos\phi_j \sigma_x+\sin\phi_j\sigma_y\rr)+\sin\vartheta_j \sigma_z\rr]\label{trot}
\ee
and similarly for any $j$.

The main trick of this approach relies on the projections on the north and south poles before and after free evolution. Only the free evolution portion which flips $\sigma_z$ survives the measurement. By using $e^{-iT_{j}H_{j}}=\cos(\omega T_j/2)-i\boldsymbol{h}_j\cdot \boldsymbol{\sigma}\sin(\omega T_j/2)$ we obtain:
\be
\ba{c}\Pi^\text{hyb}=P_{z_{\tau}}\sin(\omega T_{j+1}/2)\cos\vartheta_{j+1} \l(\cos\phi_{j+1} \sigma_x+\sin\phi_{j+1}\sigma_y\rr)\vspace{2mm}\\
\times P_{\bar{z}_{\tau}}  \sin(\omega T_j/2)\cos\vartheta_j \l(\cos\phi_j \sigma_x+\sin\phi_j\sigma_y\rr)P_{z_{\tau}}
\ea
\ee
which reduces further to:
\be
\ba{c}
\Pi^\text{hyb}=\sin(\omega T_j/2)\sin(\omega T_{j+1}/2) \cos\vartheta_{j+1}\cos\vartheta_{j} P_{z_{\tau}} \vspace{2mm}\\ \times \l[\l(\cos\phi_{j+1}\cos\phi_j +\sin\phi_{j+1}\sin\phi_j\rr) \rr.\vspace{2mm}\\\l.+i\sigma_z \l(\cos\phi_{j+1}\sin\phi_j-\sin\phi_{j+1}\cos\phi_j \rr)\rr]
\ea
\ee
And finally:
\be
\ba{c}
\Pi^\text{hyb}=p_{j,j+1} P_{z_{\tau}}  e^{i\tau_z \l(\phi_{j+1}-\phi_j\rr)}, \label{eq:hybfinal}
\ea
\ee
with
\be
p_{j,j+1}= \sin(\omega T_j/2)\sin(\omega T_{j+1}/2) \cos\vartheta_{j+1}\cos\vartheta_{j}
\ee
where we used that $P_{z_{\tau}}$ is projecting onto $\tau_z\sigma_z=-1$. So, indeed, we obtain that the state of the qubit only enters as a phase shift  $\tau_z(\phi_{j+1}-\phi_j)$. We therefore observe that the hybrid protocol allows to implement the same projectors as in a measurement-only approach [see Eq.~\eqref{eq:ns_projector}] but with the key advantage of an increased success probability. This enables an implementation of the full Chebyshev protocol as discussed in Sec.~\ref{sec:ns_measurement} with a much higher probability which supports a reasonable numbers of turning points (see simulations in Sec.~\ref{sec:numerics}).

As mentioned above, no fine tuning is required to obtain the precise phase rotation. Errors in $T_j$ and the outer coupling ($\vartheta_j\neq 0$) only suppress the probability $p_{j,j+1}^2$ of obtaining the correct measurement outcomes (i.e. alternating the distribution of outcomes of $i\gamma_0\gamma_z=\pm 1$). If $\vartheta_{j,j+1}$ is small, as expected, and $T_{j,j+1}$ could be tuned close to their required values, than this probability will be close to 1. Since we know the measurement outcomes a success probability $<1$ does not affect the fidelity of the implemented gate, it only increases the waiting time until one achieves a run with all the desired measurement outcomes.

Note that since the measurement-only and the hybrid scheme implement essentially the same projectors \eqref{eq:ns_projector} and \eqref{eq:hybfinal} the hybrid scheme is also robust with respect to unintended measurements that (partially) project the system to an eigenstate of the Hamiltonian $H_j$. The effect of the latter will only manifest in a change of the success probability which eventually reaches Eq.~\eqref{eq:ns_projector} in the limit of strong measurements. We will refer to such partial measurements in the eigenbasis of the Hamiltonian as parallel dissipation as they arise from a system-environment coupling $\propto H_j$. Note that measurements (or dissipation) that act along a vector perpendicular to $H_j$ will lead to remaining decoherence. This is, in fact, a drawback of all proposed geometric implementations of $\pi/8$ gates \cite{Clarke16,Karzig16,Plugge16}. The advantage of the hybrid scheme is rather that it is robust against the leading sources of dissipation, see discussion in Sec.~\ref{sec:realization}.

\section{Time-dependent fluctuations}
\label{sec:remaining}

On top of the systematic instrument error that our scheme is designed to eliminate, there are two remaining sources of error. The first is due to the afore mentioned non-parallel dissipation, and the second is due to temporal noise affecting the control of the system.

Non-parallel dissipation can emerge when the environment couples in an uneven way to the different Majorana pairs involved in the evolution. We discuss the physical origin of these unwanted terms in Sec.~\ref{sec:realization} and simulate the effect of static non-parallel dissipation in Sec.~\ref{sec:numerics} based on a Lindblad master equation formalism.

For the rest of this section we focus on estimating the effect of temporal fluctuations in the control. We start with the fluctuating Hamiltonian:
\begin{equation}
H(t)=\l(\frac{\omega}{2}\boldsymbol{h}+\boldsymbol{\delta}(t)\rr)\cdot\boldsymbol{\sigma}
\end{equation}

It is convenient to split the noise term into parallel and normal components:
\begin{equation}
\delta_{\parallel}(t)=\boldsymbol{\delta}(t)\cdot \boldsymbol{h}\ \ \ \ \  \delta_{\perp}(t)=\boldsymbol{\delta}(t)\cdot \boldsymbol{h}_\perp\,,
\end{equation}
with $\boldsymbol{h}_\perp=(0,0,1)\times\boldsymbol{h}$.
We now derive the effect of $\boldsymbol{\delta}(t)$ perturbatively. To first order in $\delta$ the time evolution operator in the interaction picture takes the form
\begin{eqnarray}
U_\delta &=& 1-i \int_0^T \!\!\! dt\, e^{i \frac{\omega}{2}\boldsymbol{h}\cdot \boldsymbol{\sigma}t}(\delta_\parallel(t)\boldsymbol{h}+\delta_{\perp}(t)\boldsymbol{h}_\perp)\boldsymbol{\sigma} e^{-i \frac{\omega}{2}\boldsymbol{h}\cdot \boldsymbol{\sigma}t}\ \ \ \ \ \\
&=& 1-iT \bar{\delta}_\parallel \boldsymbol{h}\cdot\boldsymbol{\sigma} -i T\delta_\perp^c \boldsymbol{h}_\perp\cdot\boldsymbol{\sigma} +i T \delta_\perp^s \sigma_z\,,\label{eq:U_delta}
\end{eqnarray}
where we defined
\begin{eqnarray}
\bar{\delta}_\parallel&=&\frac{1}{T}\int_0^T dt \delta_\parallel(t) \\
\delta_{\perp}^{c}&=&\frac{1}{T}\intt_0^T dt \delta_{\perp}(t)\cos(\pi t/T) \\   \delta_{\perp}^{s}(t)&=&\frac{1}{T}\intt_0^T dt \delta_{\perp}(t)\sin(\pi t/T) ,\,\,
\end{eqnarray}
and assumed $T\approx \pi/\omega$.

The effect of the various terms is as follows. The parallel fluctuations commute with the Hamiltonian at all times. Their main effect is an over- or under-rotation of the state quantified by the average parallel noise $\bar{\delta}_\parallel$. This only leads, in the same way as a timing error, to a slight reduction of the success probability. In Eq.~\eqref{eq:U_delta} this manifests as Pauli $\sigma_x$ and $\sigma_y$ operators that flip the state away from the south pole. Similarly, the term $\delta_\perp^c$ also only involves $\sigma_x$ and $\sigma_y$ operators. The latter is due to perpendicular fluctuations that are antisymmetric in time over the time interval $T$. The combined reduction of the success probability appears only at second order in the noise and is $\approx T^2\big( \bar{\delta}_\parallel^2+(\delta_{\perp}^c)^2\big)$.

The last term in Eq.~\eqref{eq:U_delta} leads to a phase error appearing at first order in $\delta_{\perp}^s$ by identifying
\begin{equation}
1+iT\delta_{\perp}^s\sigma_z \approx e^{iT\delta_{\perp}^s \sigma_z}.
\end{equation}
Note that since $\sin(\pi t/T)$ is positive over the entire integration window, $\delta_\perp^s$ is due to fluctuations in perpendicular direction that have a non-vanishing average over time $T$. On the other hand we expect that our cancellation scheme that involves multiple back and forth sweeps effectively cancels time dependent fluctuations slower that the time of the sweeps. The limiting time of a sweep is most likely the measurement time $T_M$ that has to pass before the state can be swept back to the north pole. To include this effect phenomenologically we calculate the variance of the phase error in terms of the noise spectral function which we cut off at frequencies smaller than $1/T_M$. Starting with
\begin{equation}
\left\langle (\delta_\perp^{s})^2 \right\rangle=\int_0^T \frac{dt dt'}{T^2} \langle \delta_\perp(t)\delta_\perp(t')\rangle \sin(\pi t/T)\sin(\pi t'/T),
\end{equation}
we introduce the spectral function of the perpendicular noise as $S_\perp(\omega)=\int dt e^{i\omega t} \langle \delta_\perp(t) \delta_\perp(0)\rangle$ which allows to rewrite the phase error variance as
\begin{equation}
\left\langle (\delta_\perp^{s})^2 \right\rangle= \int_{1/T_M}^\infty d\omega S_\perp(\omega) W(\omega T),
\end{equation}
in terms of the window function
\begin{eqnarray}
W(x)&=&\frac{1}{2\pi}\left|\int_0^1 d\tau e^{-i x \tau}\sin(\pi \tau)\right|^2 \\
&=&\frac{\pi}{(x^2-\pi^2)^2}(1+\cos x)\,.
\end{eqnarray}
Since the window function strongly decays as $1/(\omega T)^4$ for large $\omega T$ it effectively cuts off the frequency integral at $\omega \approx 1/T$.

In summary we find that only very specific time-dependent noise leads to an additional phase error. (1) The noise has to lead to perpendicular fluctuations that move the direction of the applied Hamiltonian. (2) Only the noise components within a frequency window $[1/T_M,1/T]$ are contributing. Noise much faster than the evolution time $T$ is simply averaged out during the evolution. Noise slower than the measurement time $T_M$ is canceled by the universal decoupling scheme.

\section{Numerics}
\label{sec:numerics}

In this section we study the performance of the hybrid measurement scheme. We model both, the free evolution described by the Hamiltonian $H=\omega \boldsymbol{h_{\phi_h,\vartheta_h}}\cdot\boldsymbol{\sigma}$/2 similar to that of Eq.~\eqref{rot} and the effect of the measurements. The measurements corresponding to the north and south pole can be implemented in a topologically protected way. We, therefore, describe them by the projectors $P_{z_\tau}$ and $P_{\bar{z}_\tau}$ respectively. The reduction of the trace norm of the density matrix of the system then quantifies the success probability of finding the measurement outcomes corresponding to the projectors.

For the evolution of the qubit between the poles, the environment could measure the state of the ancillary qubit which leads to decoherence. Similarly, the unprotected measurements along the equator that are part of a measurement-only implementation can also be modeled by decoherence since we do not require the knowledge of their measurement outcomes. This allows us to describe the hybrid protocol with dissipation and the measurement-only protocol on the same footing in terms of dephasing due to environmental noise along the measurement axis $\boldsymbol{l}(\phi_l,\vartheta_l)\cdot \boldsymbol{\sigma}$, where $\boldsymbol{l}(\phi_l,\vartheta_l)=\{\cos(\vartheta_l)\cos(\phi_l),\cos{\vartheta_l}\sin{\phi_l},-\sin(\vartheta_l)\}$. The time evolution of the density matrix can then be cast in form of a Lindblad master equation,
\begin{equation}
 \dot{\rho}=-i[H,\rho]+ L\rho L^\dagger - \frac{1}{2}\{L^\dagger L,\rho\},
 \label{eq:master}
\end{equation}
with $L=\sqrt{\Gamma/2} \boldsymbol{l}\cdot\boldsymbol{\sigma}$, where $\Gamma$ is the corresponding dephasing rate. The above master equation results from a system environment coupling $H_\text{SE}= \boldsymbol{l}\cdot\boldsymbol{\sigma} \Phi/2$ after integrating out the environmental degrees of freedom which are assumed to be short-time correlated $\langle \Phi(t)\Phi(0)\rangle_\text{E}=2\Gamma \delta(t)$.
Note that the assumption of a short-time correlated environment describes the worst-case scenario for our geometric decoupling scheme. Environmental noise on time scales longer than the applied back and forth sweeps would, in fact, be efficiently canceled by the Chebyshev protocol.

We numerically calculate the time evolution of the density matrix described by Eq.~\eqref{eq:master} using standard methods \cite{Johansson12, Johansson13}.

\subsection{Single-wedge example}

\begin{figure}
	\includegraphics[width=0.85\columnwidth]{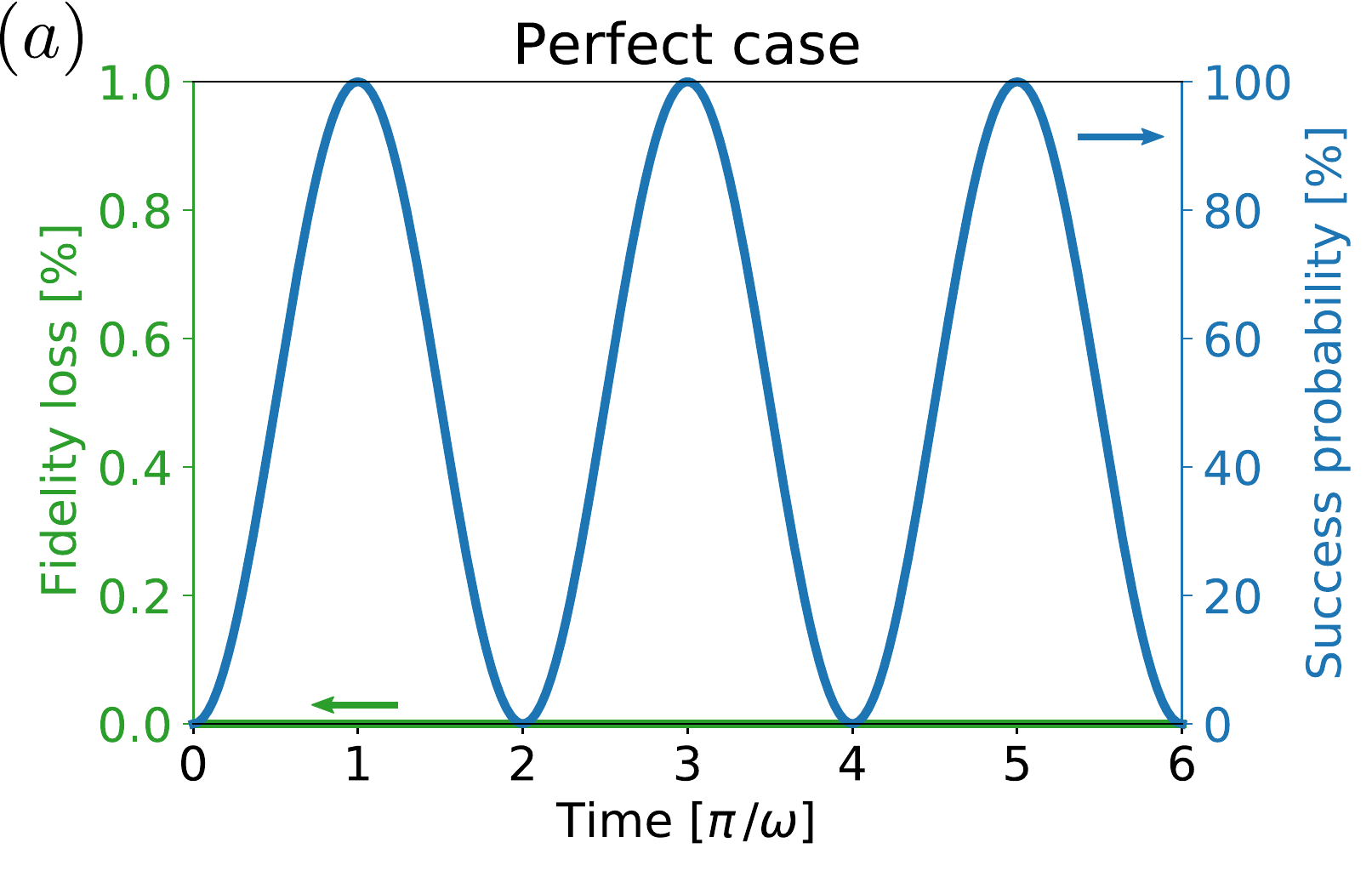}
	\includegraphics[width=0.85\columnwidth]{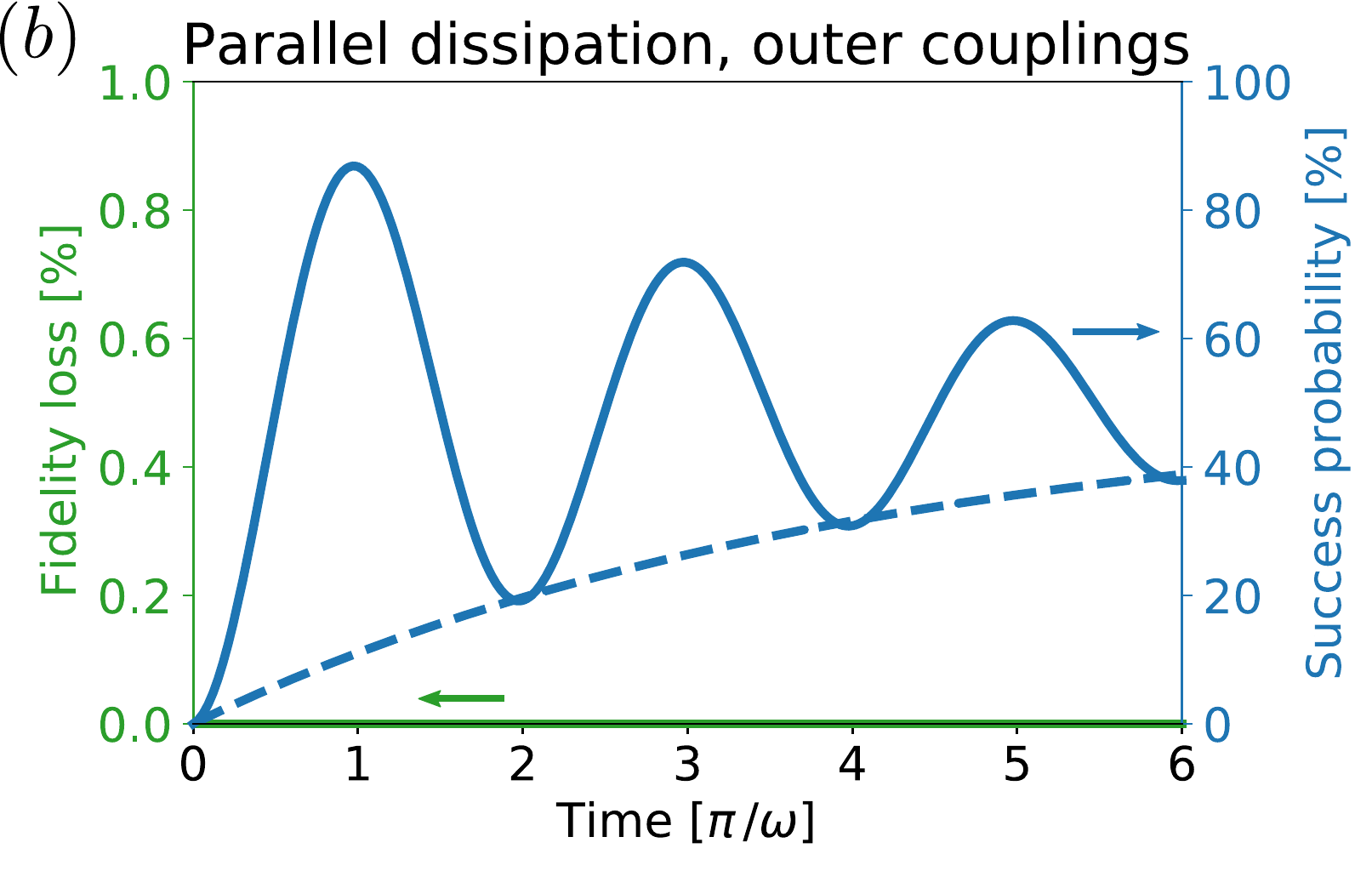}
	\includegraphics[width=0.85\columnwidth]{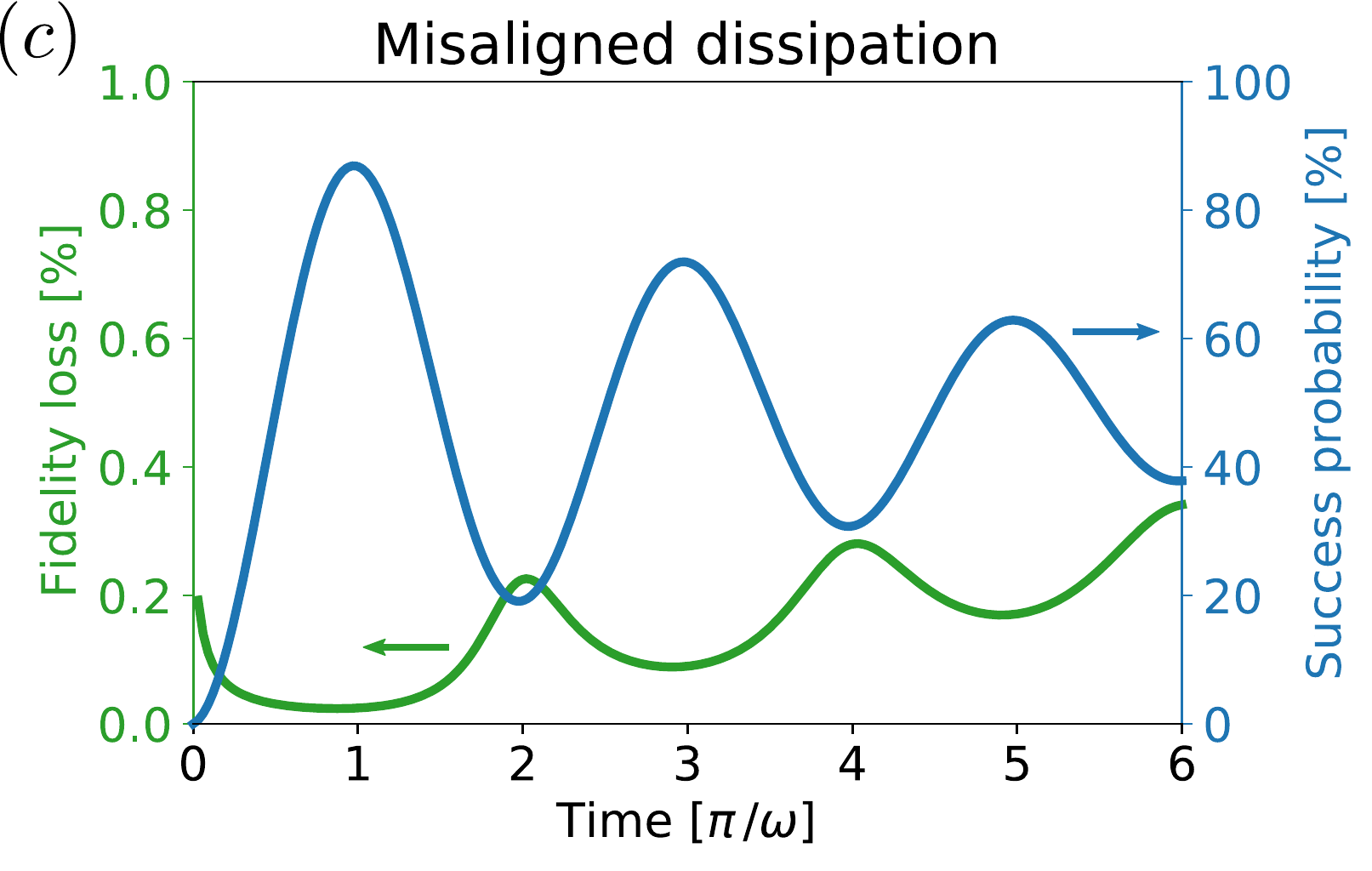}
	\caption{Simulation of a single wedge north/south sweep implementation of a $\pi/8$ gate (turning points $\phi_1=0, \phi_2=\pi/8$). In general, the timing influences both the fidelity of the gate and the success probability. The fidelity is defined as the overlap of the final state to the desired magic state given a certain set of projectors (here north, south, north) were applied. The success probability quantifies the chance of obtaining the right measurement outcomes to implement the above set of projectors. The parameters are (a) no dissipation (perfect case); (b) dissipation $\boldsymbol{l}=\boldsymbol{h}$, $\Gamma=0.5\omega/2\pi$, outer couplings $\vartheta=0.1 \pi/2$; (c) misaligned dissipation $\phi_l=1.2 \phi_2$, $\vartheta_l=1.2 \vartheta$. For comparison, the dashed line in panel (b) indicates the success probability in the case of a measurement-only procedure  $(1-\exp(-\Gamma t))/2$ purely due to dephasing (for simplicity $\vartheta=0$).}.
	\label{fig:single_wedge}
\end{figure}

The interplay of the coherent and incoherent evolution can already be demonstrated using a hybrid evolution corresponding to tracing a single wedge of the Bloch sphere, cf. Fig~\ref{fig:ns}. Specifically, consider a protocol starting with an eigenstate of $P_{z_\tau}$ (north pole), projecting onto $P_x$ (equator, $\phi=0$), followed by a projection onto $P_{\bar{z}_\tau}$ (south pole). All the measurements can be performed in a topologically protected way and desired measurement outcomes can be obtained using forced measurements. We, therefore, post-select by renormalizing the density matrix after applying the projectors. The resulting state serves as the initial state that is evolved by Eq.~\eqref{eq:master}. The measurement process is then modeled by introducing a combination of coherent ($H=\omega/2 \boldsymbol{h_{\phi_h,\vartheta_h}}\cdot\boldsymbol{\sigma}$) and incoherent ($L=\sqrt{\Gamma/2}\boldsymbol{l_{\phi_l,\vartheta_l}}\cdot\boldsymbol{\sigma}$) coupling of the Majorana modes for a time $T$. Finally, at the end of the time evolution the system is projected back to $P_{z_\tau}$ (north pole).

Let's first consider the fully coherent implementation of a $\pi/8$ gate. I.e., $\phi_h=\pi/8$, $\vartheta_h=0$ and $\Gamma=0$ (see Fig.~\ref{fig:single_wedge}a). As described by Eq.~\eqref{eq:hybfinal}. This implements a $\pi/8$ gate with certainty for $\omega T=(2n+1)\pi$. Away from this perfect timing the success probability of projecting the system to the north pole decreases. However, if the projection to the north pole is successful the resulting state is a perfect magic state with no loss of fidelity.

Next, consider adding a finite outer coupling $\vartheta_h\neq 0$ and parallel dissipation $\boldsymbol{l}=\boldsymbol{h}$, $\Gamma \neq 0$ (see Fig.~\ref{fig:single_wedge}b). The outer coupling limits the success probability to values~$<1$. Additionally, the oscillations will decay toward a $50$\% success probability with rate $\Gamma$ thus reaching the measurement-only limit for long times.

Finally, when the coherent and incoherent parts of the evolution are not aligned $\boldsymbol{h}\neq \boldsymbol{l}$ the post-selected final state of the system becomes mixed. This irreversible decoherence leads to a loss of fidelity growing over time. As expected, the highest fidelity can be achieved close to $\omega T = \pi$ (see Fig.~\ref{fig:single_wedge}c).

\begin{figure}[h]
	\includegraphics[width=0.85\columnwidth]{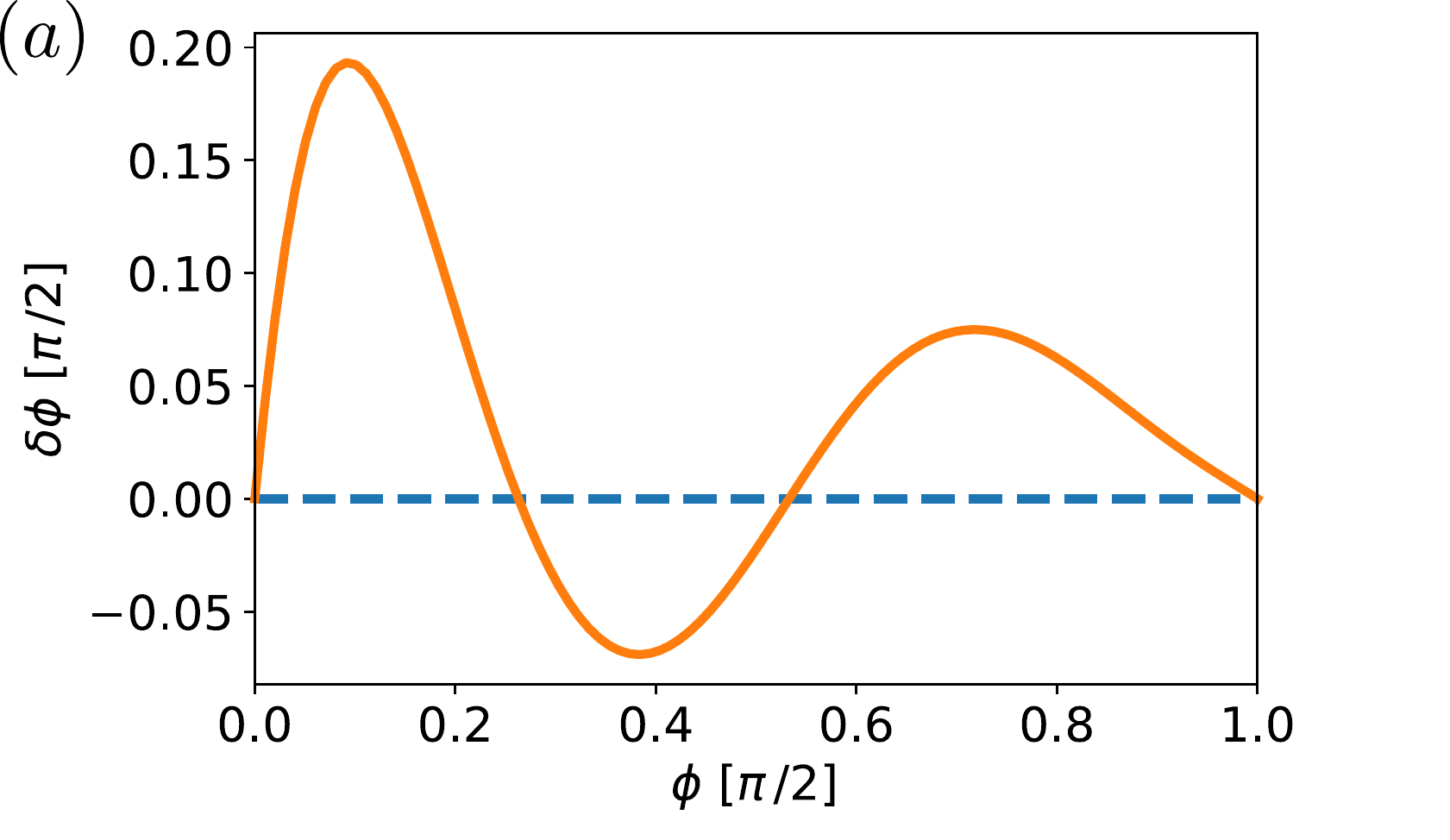}
	\includegraphics[width=0.85\columnwidth]{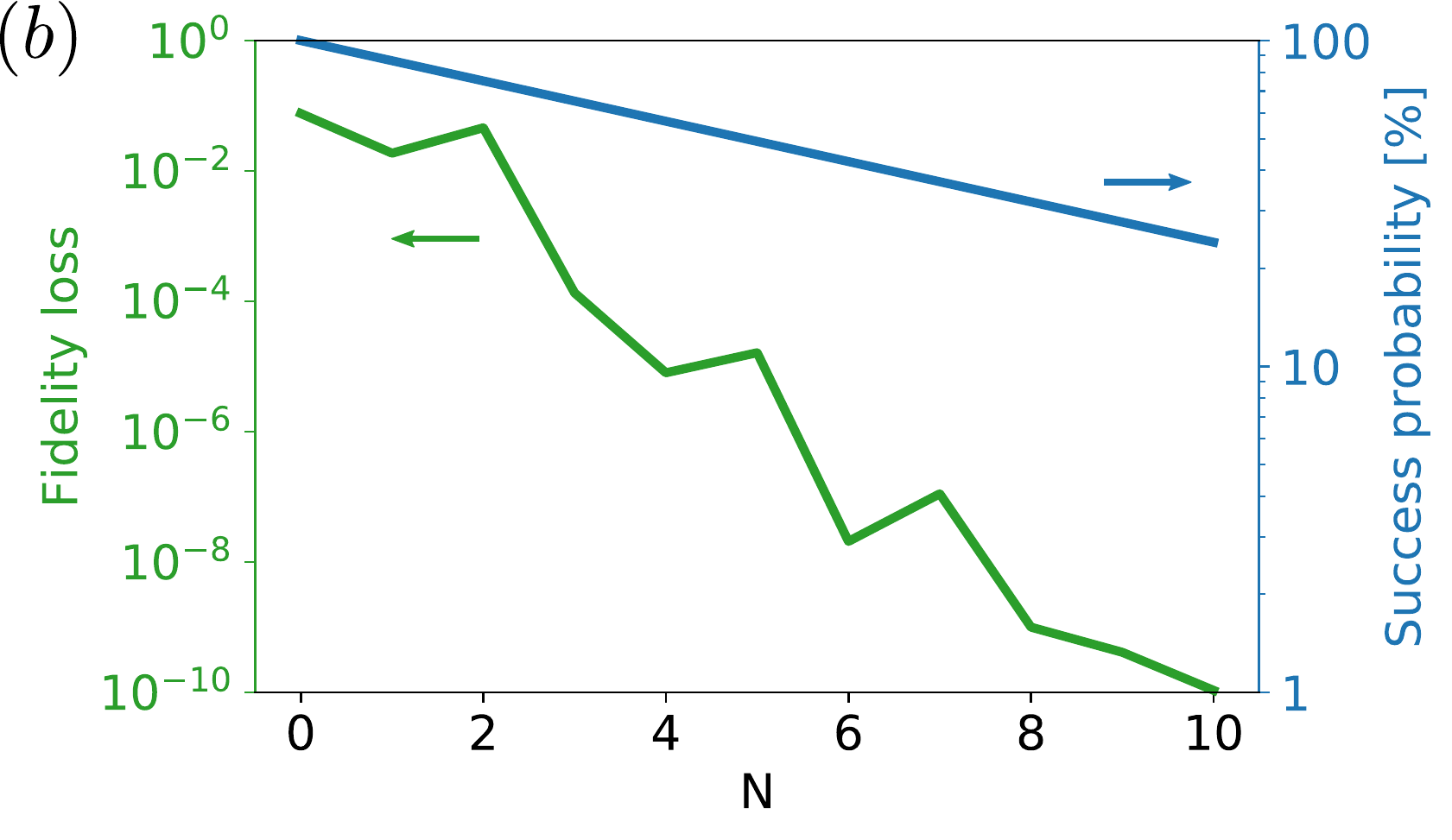}
	\includegraphics[width=0.85\columnwidth]{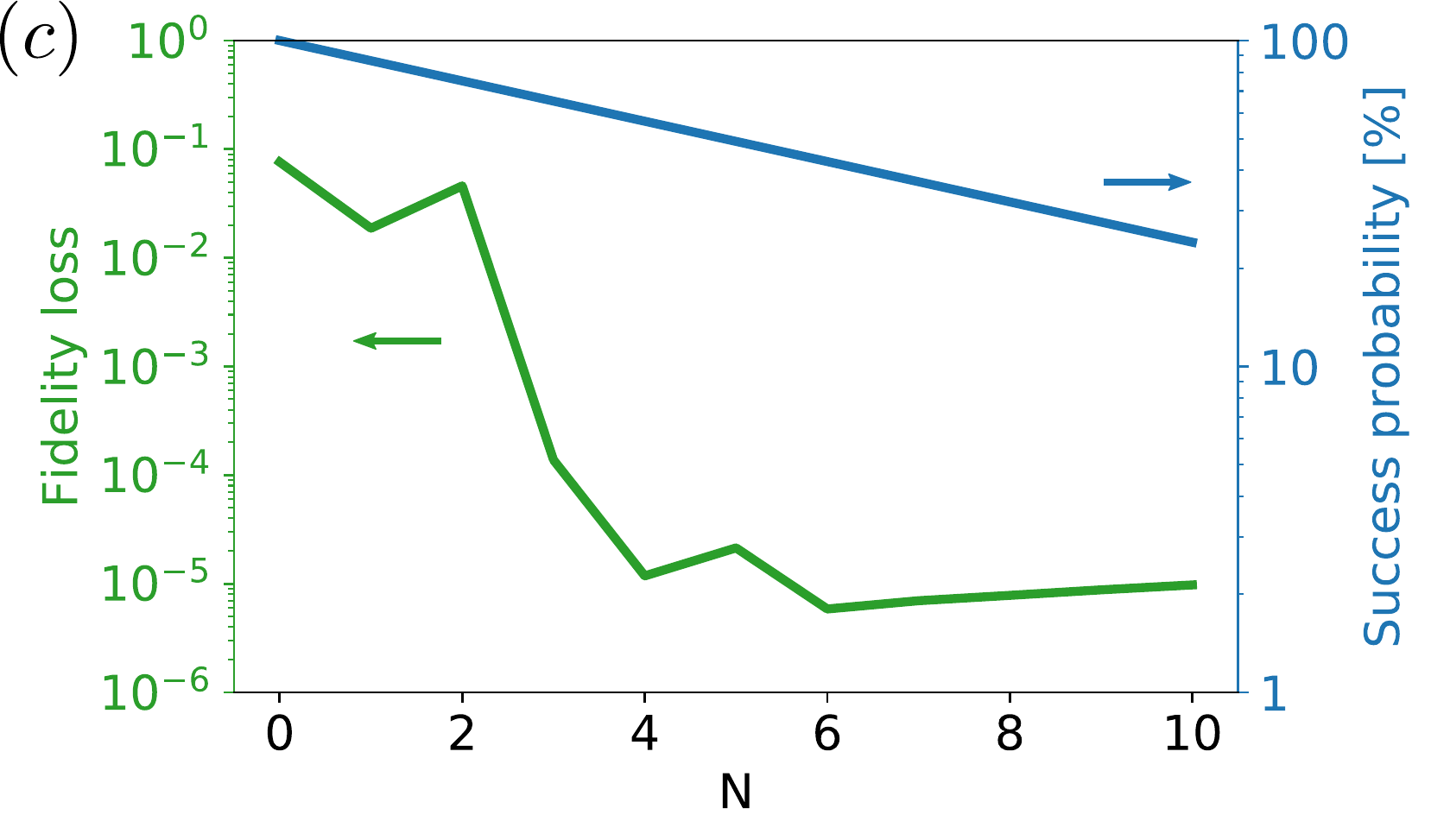}
	\caption{Simulation of the full north/south hybrid Chebyshev protocol. (a) Error function $\delta \phi=\Phi(\phi)-\phi$ relating the ideal and actual turning points. (b) Gate fidelity and success probability in terms of the number of turns $N$ with parallel dissipation $\boldsymbol{l}=\boldsymbol{h}$, $\Gamma=0.1 \omega/2\pi$, second order couplings $\vartheta_{\text{h}}=0.1 \pi/2$ and a timing offset $T=0.9 \pi/\omega$. (c) Same parameters as in (b) except for taking into account misalignment of $H$ and $L$, $\phi_{\text{l},i}=1.01\phi_{\text{h}i}$ $\vartheta_{\text{l}}=1.01\vartheta_{\text{h}}$.}
	\label{fig:full_chebyshev}
\end{figure}

\subsection{Full Chebyshev protocol}

The concepts discussed for the single-wedge example can readily be extended to simulate the full north/south hybrid Chebyshev protocol. We implement the following procedure
\begin{equation}
\rho_\text{final}= \left(\prod_{i=1}^{2N}\mathcal{P}_{z_\tau i}\, \mathcal{U}(T,\phi_{\text{h}i},\vartheta_{\text{h}i},\phi_{\text{l}i},\vartheta_{\text{l}i})\right) \rho_0,
\end{equation}
where $\rho_0$ is the initial density matrix, assumed to be an eigenstate of $P_{z_\tau}$. The superoperators $\mathcal{U}$ and $\mathcal{P}_{z_\tau i}$  implement the time evolution of Eq.~\eqref{eq:master} with fixed Hamiltonian $H_i$ and Lindblad operators $L_i$ over the time $T$ and the projections to the north ($P_{z_\tau}$, even $i$) and south ($P_{\bar{z}_\tau}$, odd $i$) pole, respectively. Ideally the angles $\phi_{\text{h}i}=\phi_{\text{l}i}=\phi_i^N
$ with $\phi_i^N$ being the Chebyshev angles extracted from solving Eq.~\eqref{eq:phin} with $\alpha=\pi/16$ \footnote{Note that in order to obtain a specific phase gate for the north/south sweep protocol $\alpha$ is smaller by a factor of 2 compared to the north/equator/north protocol discussed in Sec.~\ref{sec:GD}.}. In practice, the control and measurement apparatus of the experiment will implement different angles $\Phi_{\text{h}i}$ and $\Phi_{\text{l}i}$. To compare with the results from the original proposal of Ref.~\cite{Karzig16}, we implement a similar smooth error function $\Phi_i=\Phi(\phi_i)$ (see Fig.~\ref{fig:full_chebyshev}a).

We find that increasing the number of sweeps in the hybrid protocol proposed in Sec.~\ref{sec:hybrid} yields a similarly strong suppression of errors due to finite smooth tuning errors $\delta\phi =\Phi(\phi)-\phi$ (see Fig.~\ref{fig:full_chebyshev}b) as in the ideal case of Ref.~\cite{Karzig16}. Note, however, that the hybrid protocol does take into account finite second order couplings ($\vartheta \neq 0$) without the need of an additional echo cancellation. Moreover, the scheme is robust with respect to the most likely source of dissipation $\boldsymbol{l}=\boldsymbol{h}$ (see Sec.~\ref{sec:realization}) which would prevent a successful echo. The only downside of the hybrid approach is a success probability smaller than 1. While both the error and the success probability decay exponentially with the number of sweeps, Fig.~\ref{fig:full_chebyshev}b emphasizes that they do so with very different slopes. By optimizing the experimental accuracy, the slope of the success probability decay can, in principle, be made arbitrarily small. We used conservative estimates of $10 \%$ timing error and dissipation contribution which still yield an appreciable success probability $\sim 25 \%$ for 10 sweeps, which essentially eliminate the smooth tuning errors.

Similar to the single wedge case, misaligned dissipation $\boldsymbol{l}\neq \boldsymbol{h}$ will lead to irreversible decoherence thus limiting the maximal correction capability of the protocol. This is demonstrated in Fig.~\ref{fig:full_chebyshev}c where a misalignment of 1\% (i.e. $\phi_{\text{l},i}=1.01 \phi_{\text{h},i}$) lets the protocol saturate at an error of $\sim 10^{-5}$.

\section{Realization in hexon and tetron geometries}
\label{sec:realization}

Let us next discuss the specific implementation of the measurements and the Hamiltonian terms required for the presented hybrid protocol. We will focus on recently proposed scalable platforms for topological quantum computation composed of islands containing 4 or 6 MZMs \cite{Plugge17,Karzig17}.

The basic measurement process of projecting onto a fixed parity of a pair of MZMs (e.g., $\gamma_0\gamma_z$) is described in detail in Ref.~\cite{Karzig17}. The main idea is to connect a quantum dot to the two to-be-measured MZMs on the island. The energy levels and charge of the quantum dot will then be renormalized via co-tunneling processes with electrons entering and leaving the island through the two MZMs. In particular, there will be a parity-dependent contribution for tunneling loops that involve both MZMs. This allows to read out the parity by sensitive charge measurements of the quantum dot.

The effect of coupling quantum dots to certain MZMs can be seen from two different points of view. On the one hand, the quantum dot properties will be renormalized in a parity dependent way. On the other hand, from the point of view of the low energy Majorana degrees of freedom, the presence of the quantum dot introduces an effective coupling Hamiltonian between the MZMs. Denoting the tunnel amplitudes of the quantum dot to the MZMs by $t_0, t_ze^{i\varphi_z}$, respectively, the second order coupling Hamiltonian takes the form
\begin{equation}
H_\text{hyb}=2i\gamma_0\gamma_z\frac{t_0 t_z}{\varepsilon}\sin(\varphi_z),
\label{eq:2MZM_hyb}
\end{equation}
in terms of the detuning $\varepsilon$ of the quantum-dot/island system from degeneracy. Charge measurements of the quantum dot can then be viewed as fluctuations in $\varepsilon$ due to a coupling to the measurement apparatus. In particular, $\varepsilon$ is proportional to the gate voltage $V_g$ of the dot. The charge of the quantum dot can be obtained by the derivative of the ground state energy with respect to $V_g$. The coupling of the charge to the environment is therefore equivalent to the first order expansion of environment-induced fluctuations in $\varepsilon$ \cite{Knapp18}. Depending on whether the information of the parity state can be extracted from the environment, the fluctuations either act as a measurement or as an uncontrolled dephasing process (e.g. charge noise). When only a single pair of MZMs is coupled to external quantum dots, environmental fluctuations can only act diagonally in the parity basis of the MZM pair. In the example of Eq.~\eqref{eq:2MZM_hyb} the Lindblad operators can only be proportional to $i\gamma_0\gamma_z$ (parallel dissipation) which does not lead to any unwanted decoherence and, therefore, protects the projection into a pure state when performing a strong measurement (see Sec.~\ref{sec:numerics}).

\begin{figure}
	\includegraphics[width=.9\columnwidth]{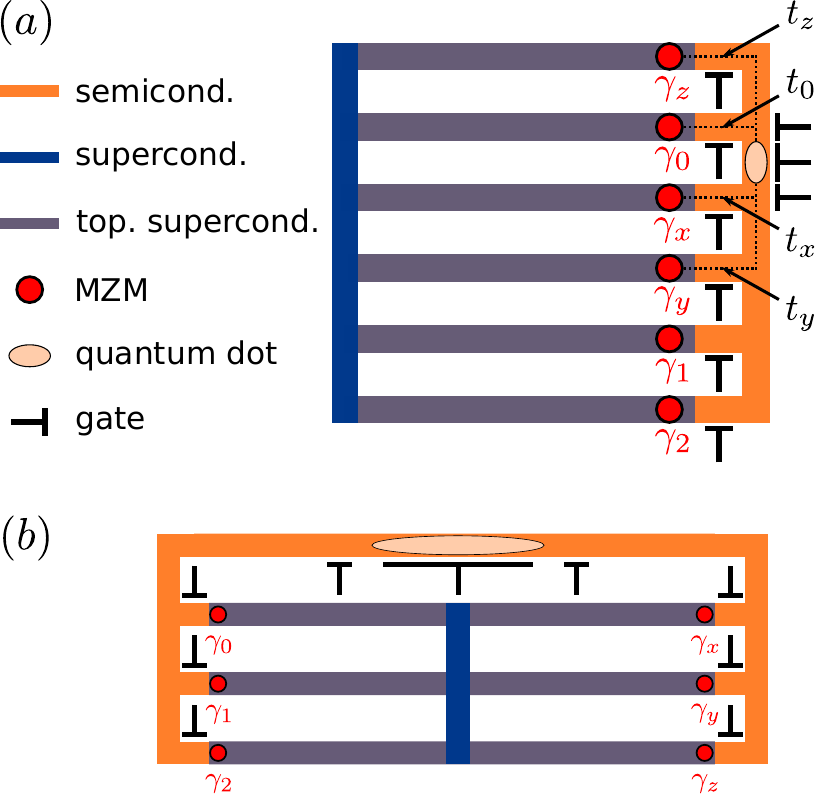}
	\caption{Experimental realization of the measurements and Hamiltonian terms required for the proposed protocols. $(a)$ One-sided hexon implementation. Finite tunnel amplitudes are denoted by dotted lines. Their strengths can be controlled by cutter gates. $(b)$ Two-sided hexon implementation which is beneficial in the presence of an approximate BDI symmetry. The quantum dot is coupled to the same $\gamma_i$ as in $(a)$ (couplings not shown).}
	\label{fig:realization}
\end{figure}

Consider now a generalization of Eq.~\eqref{eq:2MZM_hyb} to the case of 3 MZMs ($\gamma_0,\gamma_x,\gamma_y$) coupled to the quantum dot (see Fig.~\ref{fig:realization}). Defining all tunneling amplitude phases, $\varphi_x,\varphi_y$, relative to the $t_0$ coupling, yields an effective Hamiltonian of the form of Eq.~\eqref{eq:outer}, with
\begin{eqnarray}
\frac{\omega}{2} \cos\vartheta \cos\phi &=&2\frac{t_0t_x}{\varepsilon}\sin(\varphi_x) \label{eq:realization_1} \\
\frac{\omega}{2} \cos\vartheta \sin\phi &=&2\frac{t_0t_y}{\varepsilon}\sin(\varphi_y)\\
\frac{\omega}{2} \sin\vartheta &=&2\frac{t_xt_y}{\varepsilon}\sin(\varphi_y-\varphi_x)\,.
\label{eq:realization_3}
\end{eqnarray}
Changing the tunnel amplitudes therefore gives a way of tuning the direction of the Hamiltonian along the equator by adjusting the ratio of $t_x/t_y$. In physical parameters, the tunneling amplitudes are given by $t_i=g_i\sqrt{\Delta_0 \delta_\text{QD}}/2\pi$, where $g_i<1$ is the dimensionless conductance, $\Delta_0$ the topological gap, and $\delta_\text{QD}$ the level spacing of the quantum dot \cite{Heck16}. The unwanted outer couplings can be kept small in the limit $t_0\gg t_x,t_y$. Moreover, it is possible to fine-tune the phases such that $\varphi_x=\varphi_y=\pi/2$, for example by threading an appropriate flux through the tunneling loops.

If the system obeys (at least approximate) a BDI symmetry, the tunneling amplitudes are either fully real or  fully imaginary corresponding to phases $\varphi=0,\pi/2$. In a setup with parallel one-dimensional topological superconductors all the MZMs on the same side of the wire would then exhibit the same phases, thus eliminating the effective 2 MZM tunneling terms ($\varphi_x=\varphi_y=0$) in the setup Fig.~\ref{fig:realization}a  \cite{Alicea11,Chew17}. In that case one could use a modified setup (Fig.~\ref{fig:realization}b) where $\gamma_0$ is on the opposite side of $\gamma_x,\gamma_y,\gamma_z$. This fixes all phases $\varphi_{x,y,z}=\pi/2$ thus maximizing the wanted couplings and eliminating the unwanted couplings. In general, we expect BDI symmetry to be broken which makes the one-sided hexon setup (Fig.~\ref{fig:realization}a) more convenient since it minimizes the distance of the quantum dot the the MZMs.

We can qualitatively estimate the effect of dissipation by considering environmental-induced fluctuations of the parameters \eqref{eq:realization_1}-\eqref{eq:realization_3}. Charge measurements and charge noise acting on the quantum-dot/island-dipole \cite{Knapp18} couple to $\varepsilon$ and can therefore be captured by the harmless parallel dissipation. The same applies for long wavelength noise that affects the tunneling amplitudes $t_x$ and $t_y$ uniformly. The problematic transverse dissipation (with $\boldsymbol{l} \neq \boldsymbol{h} $) discussed numerically in Sec.~\ref{sec:numerics} corresponds to relative fluctuations of $t_x$ and $t_y$ (or their corresponding phases). While it is difficult to reliably estimate the strength of the remaining transverse dissipation we emphasize that in contrast to existing proposals of Majorana-based $\pi/8$ gates, the presented scheme is robust against the most prominent noise sources.

So far we focused on the implementation in 6 MZM islands (hexons). The same concepts can be used to prepare magic states using a pair of 4 MZM islands (tetrons). The hexon protocol can be viewed as measurements and/or Hamiltonian terms corresponding to $\sigma_z\tau_z$ and combinations of $\cos(\phi)\sigma_x+\sin(\phi) \sigma_y$. Note that nothing would change for the protocol if the system would start out in a $\sigma_x$ eigenstate before the first projection to $\sigma_z\tau_z$, and be projected back to a $\sigma_x$ eigenstate at the end of the protocol. This addition makes it possible to implement the same steps in a 2 tetron setup where the $\sigma$ and $\tau$ degrees of freedom are now separated as qubits of different islands. The $\sigma$ qubit acts as an ancilla and is initialized and reset to a $\sigma_x$ eigenstate. The $\sigma_z\tau_z$ projections corresponds to a joint $ZZ$ measurement (see, e.g., \cite{Karzig17}). The remaining terms acting on the $\sigma$ degrees of freedom can be implemented in a tetron similar to Fig.~\ref{fig:realization} using an island where $\gamma_1$ and $\gamma_2$ have been removed. After the final $\sigma_z\tau_z$ projection and the resetting of the ancilla, the protocol performed a $\pi/8$ phase gate on the $\tau$ qubit.

\section{Conclusions}
\label{sec:Conclusions}

The problem of realizing a protected magic gate in a Majorana system remains a key challenge of the field. This gate was believed to require either very precise control of the Majorana couplings \cite{Sau10c} or a costly distillation process. In this manuscript we showed that a sequence of measurements and free evolution applied to four MZMs eliminates the need for fine tuning, as well as the ill effects of all low-frequency noise. The only remaining sources of error, therefore, are high-frequency fluctuations,  which make changes in the device at time scales shorter than the time it takes to complete a cycle.
We explained how our scheme could be implemented in the hexon and tetron geometries, which are the likely platform for Majorana quantum information processing.

The hybrid approach affords a dramatic simplification of the scheme proposed in Ref. \cite{Karzig16}. Indeed, the latter included an echo in the Majorana manipulation intended to cancel the residual dynamical effects due to some unavoidable couplings between the MZMs. The echo increased the vulnerability of the gate to noise acting on time scales faster than the duration of the entire decoupling scheme. The hybrid approaches manages to categorically  eliminate the effects of the unwanted couplings, and, therefore, remove the need for carrying out the echo stage.

The simplification is achieved by using well chosen measurements. In order to not only simplify, but also speed up the performance of a magic gate, and make it faster than the evolution based approach, the measurement step in the hybrid scheme must not be too slow (compared to the Majorana couplings). If we estimate the parity measurement time of two MZMs by the strength of their coupling, we can speculate that the hybrid approach would give a dramatic speed-up over the adiabatic approach of Ref. \cite{Karzig16}. The free evolution stage, and the measurement stage would, ideally, each take a time $\tau\sim \omega^{-1}$. The adiabatic algorithm, in contrast, must be carried out over times $t\gg \omega^{-1}$ to avoid excitations of the manipulated qubit.

The $\pi/8$ gate we propose guarantees protections from a wide swath of errors, but it is, nonetheless, not perfect. The high expected accuracy of the scheme, however, will in the very least make the magic state this procedure produces accurate enough to be used as a good starting point for distillation schemes \cite{Bravyi05,Bravyi12,Haah17}.

For the purpose of concreteness this paper focused exclusively on making good approximations of magic states suitable for distillation into computational quality states. But we should note that very little changes if our target is a different pure state $\Psi=(e^{-i\alpha}|0\rangle+e^{i\alpha}|1\rangle)/\sqrt{2}$ on the Bloch sphere. While the closed Chebyshev form of the turning points is lost,  little is lost from the efficiency of the describe procedures.  These more general $\Psi$ might be useful directly, without distillation, for unprotected quantum circuits of order 1,000 gates which have been suggested for reaching “quantum supremacy” \cite{Boixo18,Harrow17} or for other small circuit application such as approximating the dynamics of a spin chain \cite{Childs18}. The alternative to directly producing a general $\Psi$ is to synthesize it from Clifford gates and  magic states which has its own costs; direct production in some regimes will be the better course. Furthermore, the ability to prepare smaller angle states corresponding to $\alpha=\pi/2^k$ with $k>3$, can be used to fuel more complex distillation procedures \cite{Duclos-Cianci15} which can reduce the overhead for certain quantum gates.

\section*{Acknowledgments}
We acknowledge useful discussions with Christina Knapp and Parsa Bonderson. We are grateful for the hospitality of the Aspen Center for Physics, where part of this work was performed. GR is grateful for support from the Institute of Quantum Information and Matter, an NSF frontier center. YO acknowledges the European Research Council under the European Union’s Seventh Framework Program (FP7/2007-2013) / ERC Project MUNATOP, the DFG (CRC/Transregio 183, EI 519/7-1), and the
Israel Science Foundation and the Binational Science Foundation (BSF).

\appendix

\nocite{apsrev41Control}
\bibliographystyle{apsrev4-1}
\bibliography{refpi8}

\end{document}